# Vector field visualization with streamlines.

A. Sparavigna[1] and B. Montrucchio[2]
[1]*Physics Department, Politecnico di Torino*
[2]*Control and Computer Engineering Department, Politecnico di Torino*
*Corso Duca degli Abruzzi 24*
*Torino, I-10129, Italy*

(October 15, 2006)

We have recently developed an algorithm for vector field visualization with oriented streamlines, able to depict the flow directions everywhere in a dense vector field and the sense of the local orientations. The algorithm has useful applications in the visualization of the director field in nematic liquid crystals. Here we propose an improvement of the algorithm able to enhance the visualization of the local magnitude of the field. This new approach of the algorithm is compared with the same procedure applied to the Line Integral Convolution (LIC) visualization.

## 1. INTRODUCTION

It is important in the visualization of the vector fields to use an imaging method able to grasp the underlying nature of the physical process from which these fields arise. The visualization will then succeed in enhancing the peculiar behaviors of fields and in suggesting the evolution of real systems. Vector field visualization techniques generate images where the pixels are used to give anisotropic textures conveying direction and magnitude of the field.
The main problem of visualization is to generate expressive textures without missing important details. This is in fact the problem in descriptions with glyphs, lines or arrows, when they are too sparse in the image or too dense and then overlap with a loose of information. One of the most popular methods for vector field visualization is the Line Integral Convolution, or LIC [1], first introduced in 1993. Visualization with sparse lines and with LIC related techniques suggests the structure of the vector field as an hand-drawing of the field can do (see for instance in Fig.1, a vector field representation with sparse line). We can then consider the problem of visualizing vector fields as belonging to the research area of non-photorealistic rendering (NPR) (for a recent literature survey on this subject, see Refs.2-5).
Many extensions of LIC were proposed to improve the computation speed [6, 7], to have better LIC images [8-10], and to apply LIC to both steady and unsteady flow fields [11-13]. LIC is able to depict the flow directions everywhere in a dense vector field but not the sense of the local orientation and the magnitude of the field. We have recently developed an algorithm to visualize oriented streamlines (the TOSL algorithm), which turned out to be rather useful in the analysis of the director field near topological defects in nematic liquid crystals

[14,15]. Here, we want to discuss and compare TOSL with LIC and also show how information on local intensity variation of the field can be included or enhanced in the visualization with oriented streamlines.

## 2. STREAMLINES OF A VECTOR FIELD.

Starting from a given vector field, a first step can be the visualization of its streamline pattern. Cabral and Leedom proposed in Ref.1 the LIC (Line Integral Convolution) algorithm for producing images and animations of field streamlines. The LIC algorithm assumes as input data the vector field lying on a two dimensional Cartesian grid and a white noise map with the same size of the grid. The white noise texture is a two dimensional image in which each pixel is possessing a gray tone ranging from 0 to 255, randomly distributed in the image frame. The LIC gives an output image where the input white noise map has been ``locally blurred'' according to the considered vector field. In this way, a one-to-one correspondence is established between the grid cells in the vector field region and the pixels belonging to the input texture.

In the output image, the gray tone of each pixel at the position (x,y) of the image map is obtained with a weighted average of the pixels of the white noise image along the local streamline starting at the pixel (x,y) under consideration. The weighted average producing the output pixel intensity is determined by the 1D convolution of a filter kernel applied to the pixels intensity of the input texture along the local streamline [16], according to the following formula:

$$P_{out}(x,y) = \sum_{p \in \tau} P_{in}(p)\, h(p) \qquad (1)$$

where p is the identification number of the arbitrary cell of the grid and $\tau$ is the set of the cells along the local streamline within a set distance $\pm L/2$ from the starting point (x,y). L is the length where the convolution kernel $h(p)$ is different from zero. The convolution kernel is conveniently chosen according to the imposed resolution. $P_{in}(p)$ is the pixel intensity of the input texture at the p-cell. Note that the convolution kernel $h(p)$ may be a simple rectangular filter with unit height and with basis L [1]. The computation of the streamline actually can stop when it is not possible to find any further next grid cell, for instance, when the streamline crosses the boundary of the region where the vector field is defined or when the intensity of the field along the next grid cell vanishes.

An example of local streamline is shown in Fig.2, where a vector field representing a bidimensional vortex is depicted. The streamline starts at the center of the grid cell (x,y) and moves towards the ``positive'' and ``negative'' senses for a total length L (centered in the starting point, a path L/2 is described in both senses). The LIC algorithm can be effectively employed in a large spectrum of topics, since it can be extended to visualize 3D flows [10,11,13].

## 3. LIC, OLIC AND TOSL.

Although the LIC algorithm allows an easy and fast representation of an arbitrary vector field, it is affected by the following two drawbacks: it does not provide any information about the sense of local orientation and the magnitude variations of the vector field, and it does not provide any direct information on the local azimuth gradient. The sense of flow orientation could in principle be shown through animation; unfortunately, there are cases where only few still images are available. In order to solve these problems, the OLIC (Oriented LIC) visualization was developed [17,18]. OLIC visualization uses only discrete textures with an asymmetric convolution kernel, that is ramp-like, instead of being a simple rectangular function as in LIC. With OLIC a starting average gray tone is arbitrary chosen and its intensity is varied linearly along the streamline according to the convolution filter. An example utilizing discrete textures and gray tone varying along the streamline (OLIC-like) is shown in Fig.3 (left), compared with LIC (center). A vortex is visualized with clockwise orientation, making the assumption that the local field is directed from the dark gray-towards the white tone. The result of the approach we developed, the thick-oriented streamline algorithm (TOSL), is shown in Fig.3 (right). TOSL overcomes the drawback of LIC algorithm, which does not allow the identification of the flow sense, and the drawback of OLIC, which essentially produce information loss due to its sparse texture.

According to TOSL method [15], the local streamline with arbitrary length L is computed for each grid cell of the vector field region. A first subset of the grid cells, chosen on the region of the vector field according to the Sobol's distribution [19] is considered, such as the corresponding streamlines are centered on each cell belonging to this subset. Instead of computing a convolution, the following procedure is performed to obtain the output texture: to the first pixel of these streamline, a gray tone is assigned, chosen once for all the grid cells in a pseudo-random way. Then the gray tone is assumed to change along the streamline with an increasing rate proportional to the local amplitude of the vector field in the considered cell. This means that if the vector is a unit vector, the gray tone just increases linearly. Subsequently, the grid cells not belonging to the Sobol's distribution are treated in sequence, with the same previously described procedure to assign gray tone values. The strategy of choosing a first set of grid cells with the Sobol's distribution produces an output texture where nearest neighbor streamlines do not possess correlated gray tones. The number of the first set of pixels necessary to achieve this results turns out to be ~30 % of the total cell number.

An example of TOSL algorithm representing the same field with different streamline lengths is given in Fig. 4: the vortex cells have a vector field with an amplitude varying in the image frame. As the figure shows, if the streamlines are long the field visualization is better, but, when the streamlines are too long compared with the dimensions of the area where the field magnitude variation occurs, any information on the amplitude changes is lost.

## 4. ADDING THE FIELD STRENGHT.

In Fig.5 another example of a vector field with its strength changing from point to point in the image frame is shown with LIC and TOSL visualization. In both cases we observe that the field strength variation is lost in the representation. As previously told, TOSL algorithm is considering the local amplitude but, as it happens in Fig.5 or in other cases, this is not enough to ensure a good result in the field magnitude visualization.

If it is necessary to represent all the field characteristics (streamlines and strength), a map with the field magnitude besides the map of streamlines could be given. Of course, it is better to have all the field properties on the same image. We then propose to visualize the field strength in the following manner, by performing a modification of the gray-tone map of streamlines to include the magnitude. The procedure is rather simple. The magnitude $M(x,y)$ of the field vector is evaluated for each point $(x,y)$ of the image frame to find its highest value $M_{max}$. Then the gray-tone bitmap of the field obtained with LIC or TOSL is considered. The highest value H among the gray tones appearing in this map is evaluated (the gray tone value is ranging from 0 to 255). Then the bitmap is changed. If $P_{in}(x,y)$ is the pixel of the LIC/TOSL map, an output value is determined as:

$$P_{out}(x, y) = P_{in}(x, y) \left(\frac{M(x, y)}{M_{max}}\right)^{\alpha} \frac{255}{H} \qquad (2)$$

with a parameter α that can be adjusted to enhance the rendering of the field strength. This is in fact a rather simple filtering of the streamline map. Adjusting the α parameter and the length of streamlines, visibility and quality of the image are saved.

The new map of the two vortices depicted in Fig.5 is shown in Fig.6 for LIC and TOSL. The points with lighter gray tones correspond to regions where the field strength is high. Another example is in Fig.7: the TOSL visualization applied to vortex cells with short and long streamlines is modified according to Eq.2, to enhance the detection of the field magnitude. The geometric structure of vortex cells is more evident when the image is filtered. Alternative procedures with fixed threshold values for the gray tone maps, tested on several images, were not able to give good results as Eq.(2) provides.

5. CONCLUSION.

The Thick-oriented Stream Line (TOSL) is a dense field visualization procedure. TOSL images are able to depict the streamlines and the orientation of a flow field even within a still image. Orientation is not possible to be described with LIC. Both LIC and TOSL are not able to give information on the strength of vector field and then the map of the streamlines must be modified to represent all the field characteristics. In this paper we describe a simple procedure able to insert information on the vector field magnitude. The enhancement of LIC/TOSL we propose is a filtering of the map obtained with the field visualization, according to the vector magnitude, with an adjustable parameter. If we compare enhanced TOSL with TOSL (or enhanced LIC with LIC), it is clear the advantage of an image with a clearly visible vector magnitude. Although it is possible to show the magnitude using LIC or TOSL with local variation of the streamline length L, this approach has a negative effect on the image quality. With our procedure, the quality of the image is saved or even enhanced with the adjustable parameter in the filtering. The concept of filtering the can be extended to animations. We think that filtering streamline textures can be a useful rendering of vector fields, and the enhancements given above is a fast, simple and efficient technique for the generation of such textures.

FIGURE CAPTIONS

FIG.1: Example of a sparse visualization of a vector field.

FIG.2: A 2D vector field showing the local streamline starting at the grid cell ($x$, $y$). The length set for the filter
is 13 and the cells involved by the filter in the calculation are indicated by the broken lines.

FIG.3: From left to right: OLIC, LIC and TOSL in a vortex representation. Note that OLIC and TOSL give the sense of vortex rotation.

FIG.4: TOSL visualization of vortex cells with three different streamline lengths (L=31, 71 and 101). TOSL is considering the local amplitude of the field in the increasing rate of the gray tones along the streamline. If the streamline is long, field visualization is rather good but information on local variation of the field magnitude is hindered.

FIG.5: Two vortices are represented on the left with LIC and on the right with TOSL. The strength of the field is changing in the image frame but it is not considered with LIC representation and hardly appreciable with TOSL representation.

FIG.6: The streamline maps of Fig.5 have been subjected to filtering according with Eq.(2). The magnitude of the field is higher where the gray tone is lighter.

FIG.7: Enhanced TOSL visualization of the vortex cells (L=31 on the left and L=101 on the right). Filtering the maps of Fig.5 using Eq.2 we obtain images whit a better representation of the field magnitude.

NOTE: The reader can find images with a high resolutions at the web page staff.polito.it/ amelia.sparavigna/field-visualization.

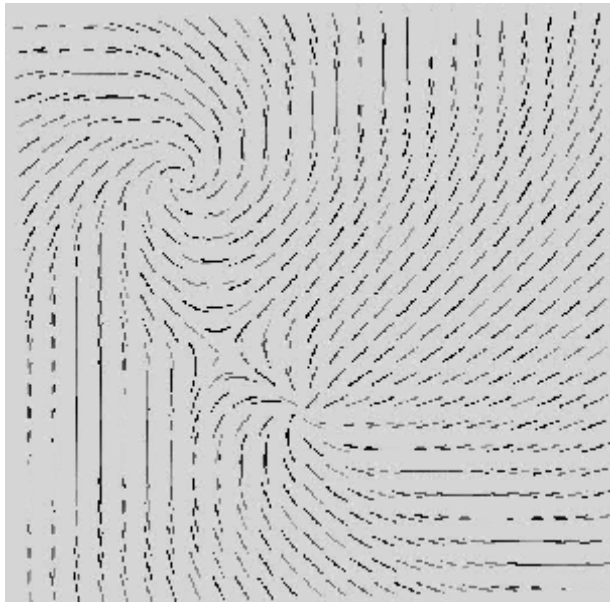

FIG.1

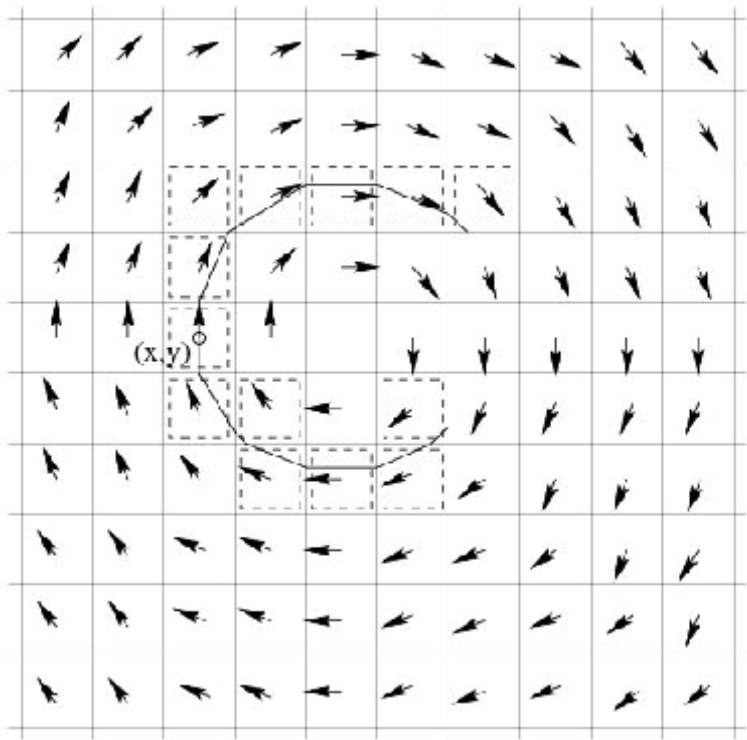

FIG.2

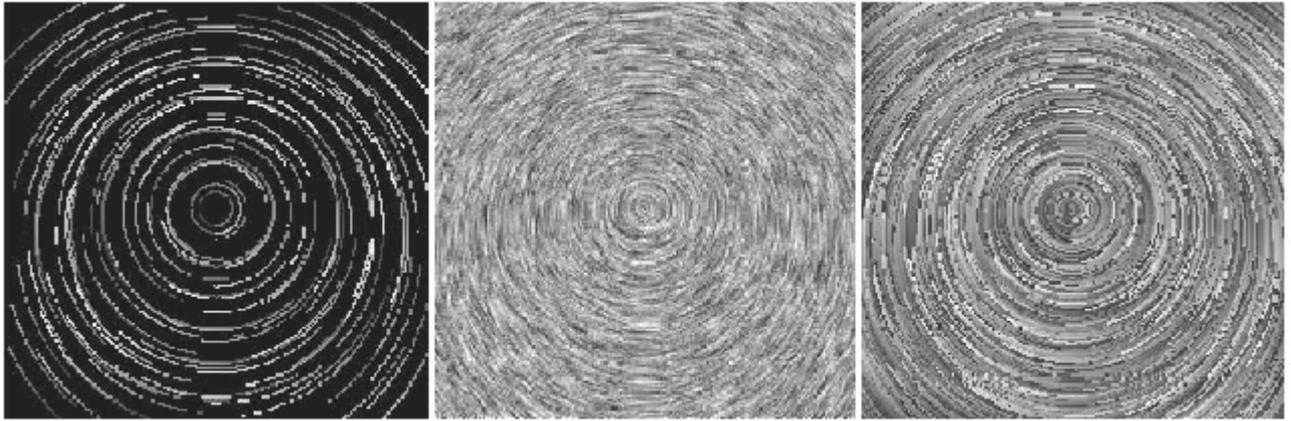

FIG.3

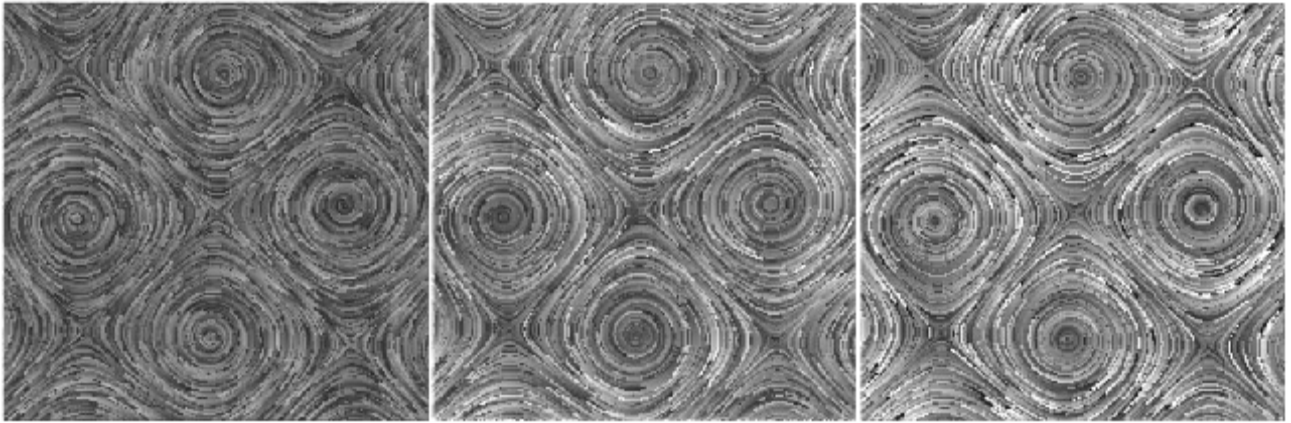

FIG.4

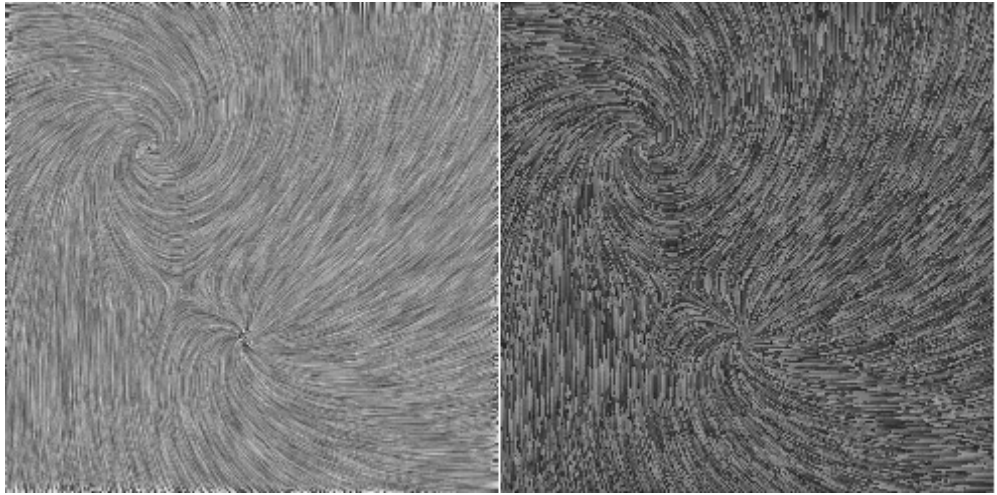

FIG.5

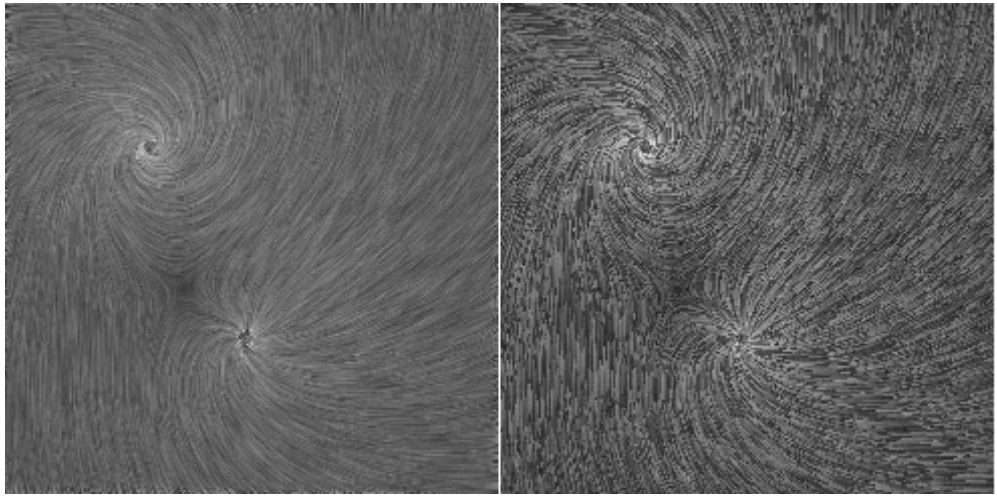

FIG.6

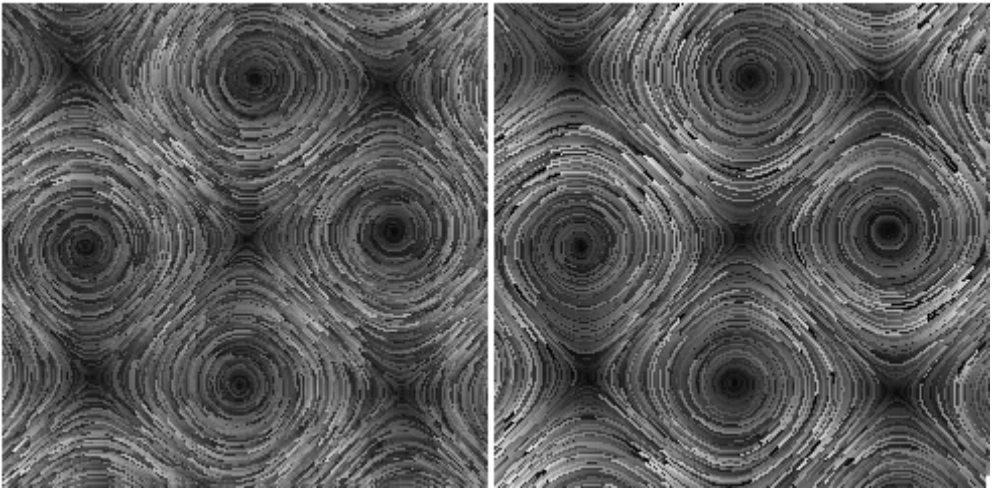

FIG. 7